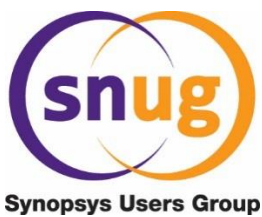

# Creation and Fixing of Lithography Hotspots with Synopsys Tools

I-Lun Tseng, Valerio Perez, Yongfu Li, Zhao Chuan Lee, Vikas Tripathi, and Jonathan Yoong Seang Ong

GLOBALFOUNDRIES Singapore Pte. Ltd.

Singapore

www.globalfoundries.com

**ABSTRACT**

*At advanced process nodes, pattern matching techniques have been used in the detection of lithography hotspots, which can affect yields of manufactured integrated circuits. Although commercial pattern matching and in-design hotspot fixing tools have been developed, engineers still need to verify that specific hotspot patterns in routed designs can indeed be detected or even repaired by software tools. Therefore, there is the need to create test cases with which targeted hotspot patterns can be generated in routed layouts by using an APR (automatic placement and routing) tool. In this paper, we propose a methodology of creating hotspot patterns in the routing space by using Synopsys tools. Also, methods for repairing hotspots during the physical design phase are presented. With the use of the proposed hotspot creation methodology, we can generate routed designs containing targeted hotspot patterns. As a result, the effectiveness of hotspot detection rules, hotspot fixing guidance rules, and relevant software tool functions can be verified.*

# Table of Contents

# Table of Figures

# 1. Introduction

A lithography hotspot pattern is a geometric layout pattern which can contain a number of polygon shapes, and the pattern is difficult to be produced correctly on silicon wafers during the photolithography stage in a semiconductor manufacturing process. At advanced process nodes (e.g., 14nm and 7nm nodes), lithography hotspots can act as major factors to cause yield loss of manufactured integrated circuits, especially under aggressive design rules [1]. To efficiently detect lithography hotspots existing in physical layouts, pattern matching techniques have been developed [2][3]. Although commercial pattern matching and in-design hotspot fixing tools have been used in the semiconductor industry, engineers still need to verify that specific hotspot patterns in routed layouts can indeed be detected or even fixed by software tools. For instance, foundries need to ensure that pattern detection and hotspot fixing guidance rules contained in their PDKs (process design kits) are correct and effective for use with EDA (electronic design automation) tools before those PDKs are released.

Since there is the need to create test cases with which targeted hotspot patterns can be generated in routed layouts by using an APR (automatic placement and routing) tool, we propose a methodology, which uses Synopsys software tools, to create hotspot patterns in routed layout designs. With the use of the proposed methodology to create test cases, we can generate routed designs containing specific hotspot patterns. As a result, the effectiveness of hotspot detection rules, hotspot fixing guidance rules, and relevant software tool functions can be verified.

This paper is organized as follows. In Section 2, three major causes of lithography hotspots occurring in physical layout designs are introduced. Also, the suitable timing for performing hotspot detection in a design flow is discussed. Section 3 describes three existing hotspot fixing methodologies; all of them can be used with Synopsys tools. In Section 4, we propose the methodology of creating lithography hotspots in the routing space with the use of Synopsys software tools, including IC Compiler. Finally, conclusions are drawn in Section 5.

# 2. Preliminary

This section describes three scenarios when undesirable lithography hotspots are induced during the physical design phase. This section also discusses the suitable timing when hotspot detection should be performed in a design flow.

## 2.1 Causes of Lithography Hotspots

In a cell-based digital design flow, each cell in a standard cell library should not contain lithography hotspots. However, undesirable hotspots can still be generated when standard cell instances are abutted. An example of a hotspot pattern is shown in Figure 1(a); the pattern contains three Metal-1 shapes, which are inside a pattern extent. As illustrated in Figure 1(b), a lithography hotspot is generated when two standard cell instances are abutted, even though there is no routed wire in the layout and each standalone cell does not contain any hotspot. Note that the red dashed rectangle in the figure denotes the location of the hotspot.

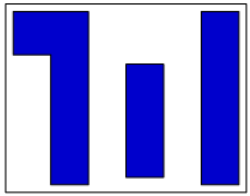

(a) Hotspot pattern containing Metal-1 shapes

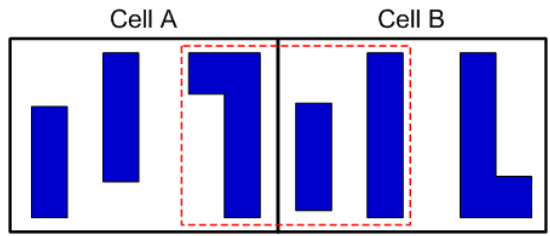

(b) Abutted standard cell instances containing a hotspot

**Figure 1. A hotspot can be generated when two standard cell instances are abutted.**

In the second scenario, undesirable hotspots are induced by routed wires only; those wires are usually created by a router. For instance, a hotspot pattern containing polygons on the Metal-2 layer and a pattern extent is illustrated in Figure 2(a), and a group of routed Metal-2 wires shown in Figure 2(b) contains a hotspot, which is marked by a red dashed rectangle.

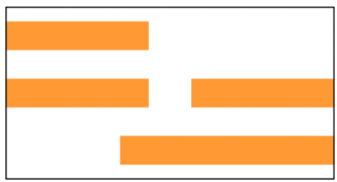

(a) Hotspot pattern containing Metal-2 shapes

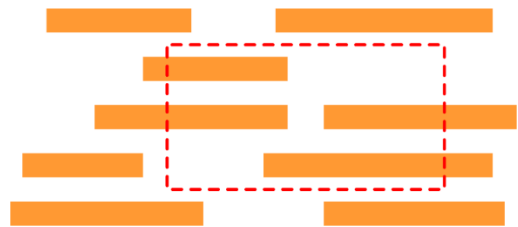

(b) Routed Metal-2 wires containing a hotspot

**Figure 2. A hotspot can be generated by routed wires.**

In the third scenario, undesirable hotspots are generated from standard cell pins together with routed wires. Figure 3(a) shows an example of a standard cell which contains four Metal-2 pins; the layout of the cell does not violate the pattern matching (PM) rule shown in Figure 2(a). However, after an instance of the cell is placed in a design and routed, extra Metal-2 shapes can be added by the router as illustrated in Figure 3(b); note that two pins of the cell instance are connected with routed Metal-2 wires, and thus a hotspot, which is defined in Figure 2(a), is generated. In the following subsection, we discuss about the suitable timing when hotspot detection should be performed in a design flow.

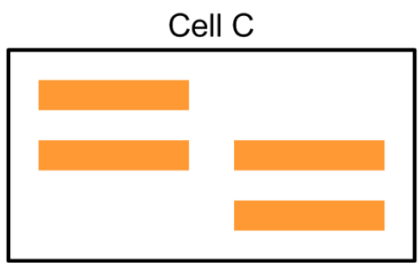

(a) A standard cell with four Metal-2 shapes

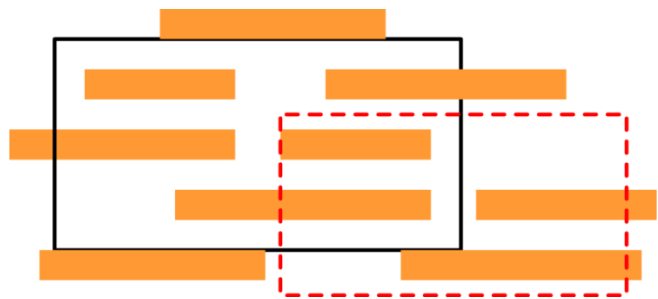

(b) A standard cell instance with routed wires

**Figure 3. A hotspot can be generated by standard cell pins together with routed wires.**

## 2.2 Detection of Lithography Hotspots in a Design Flow

Synopsys IC Validator (ICV) is a software tool which provides pattern matching (PM) functionality for detecting lithography hotspots in physical layouts. Additionally, ICV can be invoked by IC Compiler (ICC), which is an automatic placement and routing (APR) tool, for performing pattern matching during the physical design phase. Note that, in this article, we use term ICV-PM to denote the pattern matching functionality of ICV.

During the process of developing standard cells, engineers should verify that the layout of each standalone standard cell does not violate PM rules; these verification tasks, which involve the detection of lithography hotspots, can be performed by using ICV-PM. Moreover, standard cell developers should ensure that abutted standard cell instances do not induce hotspots if the cell instances can be abutted. Cell spacing constraints should be used if two abutted cell instances generate hotspots. On the other hand, if a given standard cell library cannot be guaranteed to be free from lithography hotspots whenever two cell instances are abutted, hotspot detection should be performed after the placement stage is completed. Section 3.2 presents a methodology for repairing hotspots which are induced by abutted standard cell instances.

In Section 2.1, the second and the third scenarios for the causes of hotspots are discussed. Since these two scenarios are relevant to routed wires, hotspot detection should be performed after detailed routing is completed. In Sections 3.1 and 3.3, methodologies for repairing hotspots which belong to these two scenarios are presented.

# 3. Fixing Lithography Hotspots

This section presents three methodologies for repairing lithography hotspots during the physical design phase. These methodologies include (1) rip-up and re-route, (2) cell flipping and shifting, and (3) surgical fixing. Note that it is usually desirable to fix hotspots early in a design flow.

## 3.1 Rip-up and Re-route

Rip-up and re-route is a traditional methodology of repairing lithography hotspots after performing detailed routing [4][5]. The methodology removes one or more wires causing hotspots and then performs re-routing for those removed wires. Synopsys provides an automatic DRC repair (ADR) approach, which uses ICV-PM to detect and mark hotspots, and then uses ICC to perform rip-up and re-route based on the information of those marked hotspots. However, this methodology can affect circuit timing and may require several iterations of the process, resulting in a long turnaround time. Therefore, rip-up and re-route should be used as the last resort to fix lithography hotspots.

## 3.2 Cell Flipping and Shifting

As discussed in Section 2.1, lithography hotspots can be generated when two standard cell instances are abutted and before the routing is performed. In this scenario, we can perform automatic cell flipping and/or cell shifting to repair hotspots [6]. To repair the hotspot in the example of abutted cells shown in Figure 1(b), we can flip Cell B as the layout shown in Figure 4(a). Alternatively, as illustrated in Figure 4(b), we may either shift Cell A to the left or shift Cell B to the right in order to repair the hotspot.

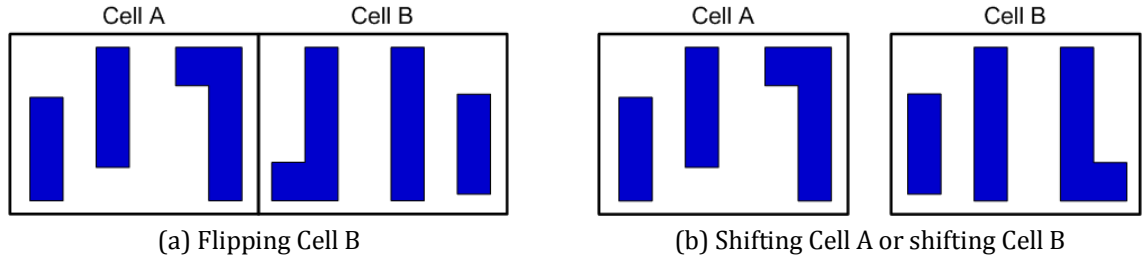

<p align="center">(a) Flipping Cell B       (b) Shifting Cell A or shifting Cell B</p>

**Figure 4. Cell flipping and cell shifting.**

### 3.3 Surgical Fixing

Surgical fixing is a methodology for repairing a hotspot by using localized modifications of geometric shapes which form the hotspot, so that the modified pattern does not contain any hotspot and does not violate design rules. Figure 5 illustrates an example of surgical fixing; Figure 5(a) shows a hotspot pattern and Figure 5(b) shows another pattern which is not a hotspot. The example indicates that the original hotspot can be repaired by adding a small rectangular shape, which is marked as a red-outlined rectangle in Figure 5(b). One advantage of surgical fixing is that the methodology has negligible impact on circuit timing. Additionally, our experiments show that surgical fixing can have high fix rates and short turnaround time.

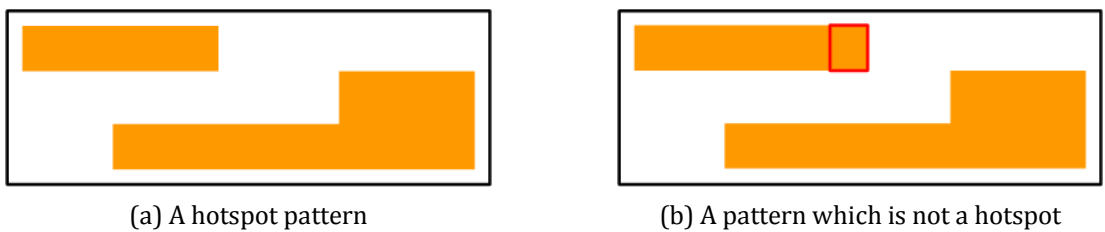

<p align="center">(a) A hotspot pattern       (b) A pattern which is not a hotspot</p>

**Figure 5. A hotspot pattern and a resultant pattern which is repaired by using surgical fixing.**

The Automatic DRC Repair (ADR) approach provided by Synopsys not only can use rip-up and re-route to repair hotspots, but also can use surgical fixing methodology to perform the repairing process [7][8]. For each detected hotspot, the ADR approach performs surgical fixing on it if relevant fixing guidance rules exist for the hotspot and the fixing does not violate design rules. Rip-up and re-route will be carried out if there is no fixing guidance for the hotspot or surgical fixing fails to repair it.

To use the surgical fixing functionality provided by Synopsys tools, fixing guidance rules must be defined. These fixing guidance rules are usually provided by foundries. GLOBALFOUNDRIES provides fixing guidance rules as part of process design kits (PDKs) for 7nm, 12nm, 14nm, 22nm, and other process nodes.

## 4. Creation of Lithography Hotspots

This section presents our hotspot creation methodology, which is based on a LEF/DEF [9] physical design flow with the use of Synopsys software tools. Hotspots created by using the methodology not only can be used as test cases for verifying the effectiveness and correctness of hotspot fixing guidance rules, but also can be used for testing hotspot detection rules and relevant software functions.

## 4.1 Overview of the Hotspot Creation Methodology

The hotspot creation flow that is currently in use at GLOBALFOUNDRIES is illustrated in Figure 6. In the flow, five types of data are needed. These types of data include (1) a standard cell library in Milkyway database format, (2) a netlist in Verilog format, (3) a standard cell placement in DEF format, (4) a technology file which is for use by ICC, and (5) a Tcl runset file which includes a set of ICC commands. Note that technology files are usually provided by foundries.

After all of the required data are prepared, we can invoke ICC to perform non-timing-driven routing, as shown in Figure 6. ICV-PM can then be used to check if the routed design contains targeted hotspots. The input data may need to be modified if the targeted hotspots are not created.

In this paper, one major goal of creating hotspots in physical layouts is to verify that hotspot fixing guidance rules are correct and effective for use with the Synopsys ADR approach in order to repair hotspots by using surgical fixing after detailed routing is completed, although the created hotspots can also be used to verify the correctness of hotspot detection rules and can be used for other purposes. Also, to enable the surgical fixing functionality provided by the ADR approach, specific shapes in a hotspot must be nets. Therefore, during the process of creating hotspots, we have to make ICC generate nets in desired locations and with desired dimensions. Thus, for a hotspot pattern containing several polygons, we first have to decide which polygons will be generated as routed nets and which polygons will be included in a standard cell as pins or intrinsic shapes of the cell. That is because the ADR approach, which involves the use of ICC, only adds or removes shapes on polygons which belong to the type of routed nets.

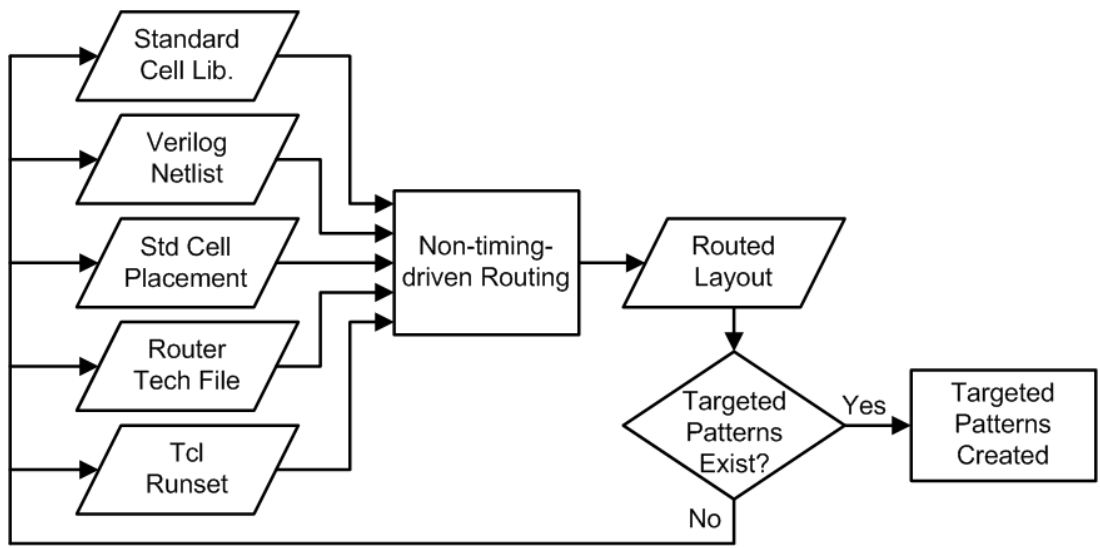

**Figure 6. Proposed hotspot pattern creation flow.**

## 4.2 Preparation of a Standard Cell for a Targeted Hotspot

To create a hotspot shown in Figure 5(a), we can design the layout of a standard cell as shown in Figure 7; the cell has three pins, which are pins A1, A2, and A3. One of the reasons for designing the standard cell layout as shown in the figure is because we plan to let pins A1 and A2 connect by a wire, so that the routed standard cell instance will form a hotspot. Moreover, we hope that the created hotspot can be repaired by surgical fixing, so that we can verify the correctness and effectiveness of the hotspot fixing guidance rules. Therefore, after we have decided which shapes will become standard cell pins, we can specify the standard cell layout in the Library Exchange Format (LEF). Figure 8 shows an example of a LEF file for a standard cell library, which contains only one standard cell whose layout is shown in Figure 7. Note that the standard cell is named `MY_GATE` in the LEF file.

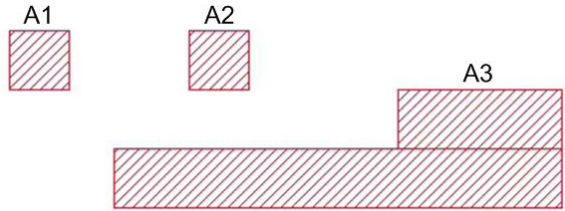

**Figure 7. Designing pins of a standard cell for the purpose of creating a hotspot.**

```
VERSION 5.8 ;                          PIN A2
BUSBITCHARS "[]" ;                       DIRECTION INPUT ;
DIVIDERCHAR "/" ;                        ANTENNAMODEL OXIDE1 ;
SITE MY_SITE                             ANTENNAGATEAREA 0.5 ;
     SIZE 0.400 BY 1.120 ;               PORT
     CLASS CORE ;                          LAYER M2 ;
     SYMMETRY Y ;                          RECT 1.490 0.460 1.530 0.500 ;
END MY_SITE                              END
MACRO MY_GATE                          END A2
  CLASS CORE ;                         PIN A3
  ORIGIN 0 0 ;                           DIRECTION INPUT ;
  SIZE 3.200 BY 1.120 ;                  ANTENNAMODEL OXIDE1 ;
  SYMMETRY X Y ;                         ANTENNAGATEAREA 0.5 ;
  SITE MY_SITE ;                         PORT
  PIN A1                                   LAYER M2 ;
    DIRECTION INPUT ;                      POLYGON 1.440 0.380 1.740 0.380
    ANTENNAMODEL OXIDE1 ;              1.740 0.460 1.630 0.460 1.630 0.420
    ANTENNAGATEAREA 0.5 ;             1.440 0.420 1.440 0.380 ;
    PORT                                   END
      LAYER M2 ;                         END A3
      RECT 1.370 0.460 1.410 0.500 ;   END MY_GATE
    END
  END A1                               END LIBRARY
```

**Figure 8. The LEF file for a standard cell library containing one standard cell.**

In order to perform routing by using IC Compiler (ICC), the input standard cell library has to be in Milkyway database format. We can use Synopsys Milkyway tool to convert a standard cell LEF file into a Milkyway database format. As shown in Figure 9(a), the conversion can be performed by using the function of "Cell Library → LEF In …", which is located on the GUI menu of the Milkyway tool. The conversion process will be invoked after the user fills in required information on the dialog box of "Read LEF", as shown in Figure 9(b), and then clicks on a button titled "OK".

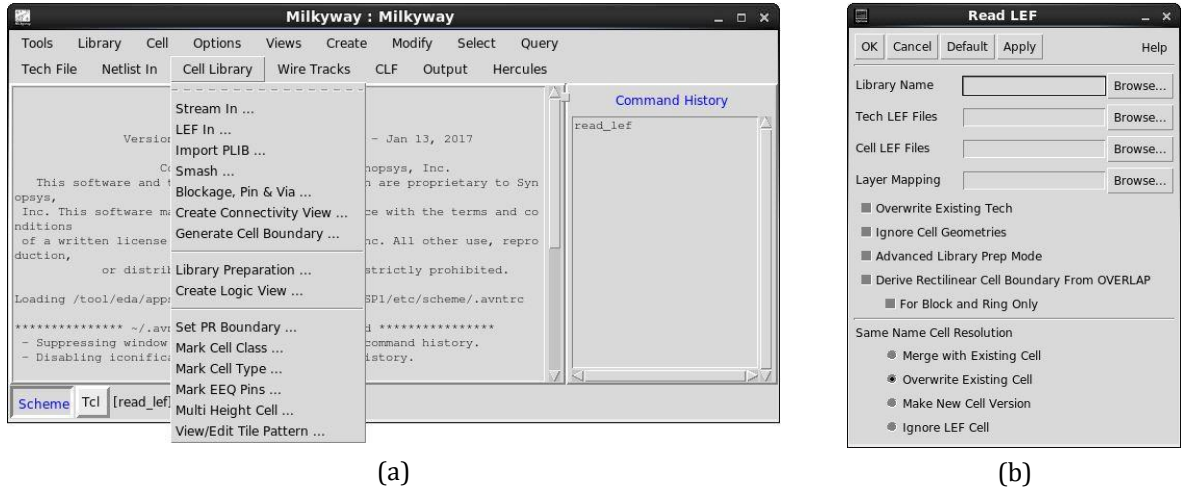

(a)                                                                 (b)

**Figure 9. Converting a cell LEF file into a Milkyway database with Synopsys Milkyway tool.**

## 4.3 Preparation of a Verilog Netlist

To connect pins A1 and A2 for an instance of the standard cell shown in Figure 7 by using ICC, we can prepare a Verilog netlist which specifies that the two pins are connected by a net. Figure 10 shows such an example of a Verilog netlist. In the netlist, the standard cell instance is named `X1`, and a net named `net1` connects the two pins of the cell instance.

```
module top ( );
    wire   net1;
    MY_GATE X1 ( .A1(net1), .A2(net1) );
endmodule
```

**Figure 10. An example of a Verilog netlist.**

## 4.4 Preparation of a DEF File

In the proposed hotspot creation methodology, the data for representing a placement of standard cell instances is needed. The placement can be specified by a DEF file, which describes the locations of standard cell instances. For the example of the standard cell LEF in Figure 8 and the Verilog netlist in Figure 10, we can prepare a DEF file as shown in Figure 11; the DEF file contains the placement for only one cell instance named `X1`.

```
VERSION 5.8 ;
DIVIDERCHAR "/" ;
BUSBITCHARS "[]" ;
DESIGN top ;
UNITS DISTANCE MICRONS 1000 ;
DIEAREA ( 0 0 ) ( 4000 2000 ) ;

COMPONENTS 2 ;
- X1 MY_GATE + PLACED ( 0 0 ) N ;
END COMPONENTS

END DESIGN
```

**Figure 11. An example of a DEF file.**

## 4.5 Preparation of an ICC Runset

To make ICC perform non-timing-driven routing, we can prepare an ICC runset in Tcl format. Figure 12 shows a template of such a runset. Note that the preferred routing direction for each routing layer may need to be specified. Also, routing tracks may need to be removed and re-created before `route_zrt_auto` is performed. Figure 13 shows the final layout after ICC performs routing for the standard cell shown in Figure 7. Note that the routed layout contains a hotspot defined in Figure 5(a).

```
set mw_design_lib   <std_cell_lib_in_Milkyway_database_format>
set techfile        <ICC_technology_file>"
set netlist_file    <Verilog_netlist_file>
set design_name     <design_name>
set def_file        <placement_DEF_file>
open_mw_lib $mw_design_lib
import_designs $netlist_file -format verilog -top $design_name -cel $design_name
set_mw_technology_file -technology $techfile $mw_design_lib
read_def $def_file
route_zrt_auto
save_mw_cel
```

**Figure 12. A template of an ICC runset for creating hotspots.**

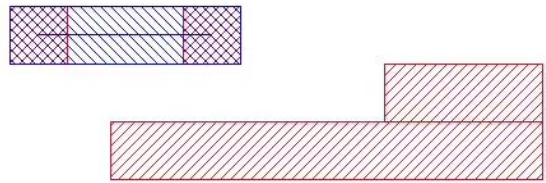

**Figure 13. The layout illustrating that a hotspot is generated after performing routing.**

## 4.6 Performing Surgical Fixing for Testing Created Hotspots

After a hotspot is created by using the proposed methodology, we can perform in-design hotspot fixing in order to verify that the created hotspot can be repaired during the physical design phase. The in-design hotspot fixing can be performed with the use of ICC, which invokes ICV to perform hotspot detection. The in-design hotspot fixing flow can be executed via running a set of ICC commands as shown in Figure 14. With proper hotspot fixing guidance rules, the created hotspot can be repaired by surgical fixing and thus the final fixed layout is shown in Figure 15. Note that the in-design fixing flow may also use the rip-up and re-route methodology to repair a hotspot if relevant fixing guidance rules do not exist or surgical fixing fails to repair the hotspot.

```
set_physical_signoff_options -default -exec_cmd icv -drc_runset
<GlobalFoundries_ICVPOP_Runset> -mapfile <layer_map_file>

report_physical_signoff_options

# Run ICV to perform hotspot detection
signoff_drc -run_dir <output_directory_#1> -ignore_child_cell_errors
-max_errors_per_rule <value> -ignore_fill_view_time_stamp

# Hotspot Fixing (may use surgical fixing or rip-up and re-route)
signoff_autofix_drc -config_file auto -init_drc_error_db
<output_directory_#2> -max_errors_per_rule <value>

# Run ICV again for hotspot detection after hotspot fixing
signoff_drc -run_dir <output_directory_#3> -ignore_child_cell_errors
-max_errors_per_rule <value> -ignore_fill_view_time_stamp
```

**Figure 14. ICC commands for performing in-design hotspot fixing.**

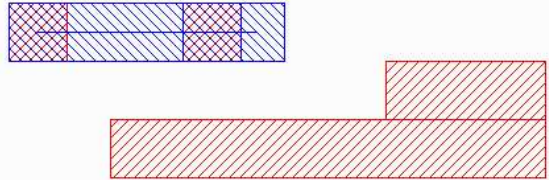

**Figure 15. The layout illustrating that a hotspot is repaired by surgical fixing.**

## 4.7 Guidelines for Creating Hotspots

Many details should be taken into consideration when the proposed hotspot creation methodology is used. A number of these details are listed below.

- Some layers in a hotspot cannot be generated by a router if those layers are not routing layers
- Preferred routing direction for each routing layer should be considered since a router usually generates a wire in the preferred direction
- To control the location and the dimensions of a routed wires, factors including routing tracks, default wire width, via dimensions, and standard cell pins should be considered
- Routing blockages can be used to prevent a router from generating wires at undesired locations
- Created hotspots should not violate design rules, otherwise the surgical fixing may not work
- It is recommended to create one or more hotspots by using one standard cell only so that placement results are unimportant
- Since one layout design can be composed of many standard cell instances, one layout design can contain many targeted hotspots

# 5. Conclusions

In this article, we have introduced causes of lithography hotspots which are induced during the physical design phase, and presented methodologies of repairing hotspots. Additionally, we have proposed a methodology for creating targeted hotspots by using Synopsys software tools, which include IC Compiler and Milkyway, so that we can verify if those created hotspots can be detected and verify if the hotspots can be repaired by using surgical fixing, rip-up and re-route, and relevant software functions. Our future work includes the development of a software system to automatically generate test cases which are used for creating targeted hotspots in the routing space.